\documentclass[letter,twocolumn, nofootinbib,showkeys,showpacs,10pt]{revtex4}
\usepackage{amssymb}
\usepackage{amsmath}
\usepackage{makeidx}
\usepackage{amsfonts}
\usepackage{graphicx,epstopdf}
\usepackage[ansinew]{inputenc}
\usepackage[usenames,dvipsnames]{pstricks}
\usepackage{subfigure}
\usepackage{pdfpages}
\usepackage{epsfig}
\usepackage{pst-grad}
\usepackage{pst-plot}
\usepackage[colorlinks,hyperindex]{hyperref}

\hypersetup{
colorlinks,
citecolor=blue,
linkcolor=black,
urlcolor=black,
}

\newcommand{\be}{\begin{equation}}
\newcommand{\ee}{\end{equation}}

\DeclareMathOperator{\sech}{sech}
\newcommand{\beq}{\begin{eqnarray}}
\newcommand{\eeq}{\end{eqnarray}}

\begin{document}

\title{Configurational Entropy in Brane-world Models: A New Approach to
Stability}
\author{R. A. C. Correa}
\email{rafael.couceiro@ufabc.edu.br}
\affiliation{  CCNH, Universidade Federal do ABC, 09210-580, Santo André, SP, Brazil}
\author{Rold\~ao da Rocha}
\email{roldao.rocha@ufabc.edu.br}
\affiliation{CMCC, Universidade Federal do ABC, 09210-580, Santo André, SP, Brazil}
\affiliation{International School for Advanced Studies (SISSA), Via Bonomea 265, 34136
Trieste, Italy}

\begin{abstract}
In this work we investigate the entropic information on thick brane-worlds
scenarios and its consequences. The brane-world entropic information is studied for the 
sine-Gordon model is and hence the brane-world entropic
information measure is shown an accurate way for providing the most suitable
values for the bulk AdS curvature. Besides, the brane-world configurational
entropy is employed to demonstrate a high organisational degree in the
structure of the system configuration, for large values of a parameter of
the sine-Gordon model but the one related to the AdS curvature. The Gleiser
and Stamatopoulos procedure is finally applied in order to achieve a precise
correlation between the energy of the system and the brane-world
configurational entropy.
\end{abstract}

\maketitle

\section{Introduction}

The 4D Universe can be regarded as a brane embedded in a higher-dimensional
bulk, in which extra dimensions can be large \cite{ADD,ant} and either compact or non-compact \cite%
{shapo,vissera,rs1,gog,Lykken}. Brane-world models have been providing new tools to understand various questions, as for
instance the solution of the gauge hierarchy problem \cite{rs1,gog,ADD}.

Recently, an widening interest has been focused on the study of thick brane
scenarios based on gravity coupled to one or more scalars in higher
dimensional space-times \cite{De_Wolfe_PRD_2000, gremm,
Cvetic1,cvt3,cvt4,koba,campos,folo,bazeia,bazeia1,ThickBrane2,nucaa,Jardim:2011gg,Liu:2011nb,Bernardini:2013tba,German:2012rv,Casana:2013fva,Cruz:2014eca,Correa:2010zg}%
. Moreover, domain walls have been used in high-energy physics to generate
thick branes in models wherein scalar fields can couple with gravity in
warped spacetime. Besides, thick branes
are well-known to avoid some inconsistencies inherent to thin brane-worlds,
as for instance the proton decay \cite{hoff1}. For some prominent reviews on domain walls and thick
branes see, e. g., Refs. \cite{Cvetic2,0904.1775}. In particular, a
model described by a single scalar field with internal structure was
proposed in Refs. \cite{bazeia1,bame}.  Analytical solutions of the Einstein
equations were achieved with a sine-Gordon potential \cite{kkar}, where a
kink, as the scalar field configuration in the system, provides the thick
brane-world as a domain wall in the bulk. Similarly, this type of
configuration has been also further explored \cite{ichi, ring}.
In addition, an analytic solution of the sine-Gordon domain wall in a 4D global
supersymmetric model was obtained \cite{sakai}, and the stability of a more
general setup was studied likewise \cite{eto}. The localization of fermions
on the brane has been accomplished in the presence of kink-fermion couplings
in the background of the sine-Gordon kink \cite{Cruz:2014eca}.

Our prominent aim here is to study the entropic information on thick
brane-worlds models, by means of the brane-world configurational entropy,
and to further explore its subsequent ramifications. In order to accomplish it,
the sine-Gordon kink shall be used, playing a prominent role on the thick
brane scenario.

This paper is organized as follows: in the next section the thick brane
generated by a single scalar field coupled to gravity shall be revisited,
with particular attention to the sine-Gordon model. In Sec.  III the
brane-world configurational entropy is going to be introduced, and we shall
calculate the entropic information for the sine-Gordon model, using the
Fourier transform of the thick brane (warped) energy density. The entropic
information measure shall be shown a successful manner for constraining the
most suitable values for the AdS curvature. In addition, the configurational
entropy is employed to evince a high organisational degree in the
configuration of the system, for large values of a parameter of
the sine-Gordon model. Furthermore, the Gleiser and Stamatopoulos procedure
is going to be applied to obtain a correlation between the brane-world
configurational entropy and the energy of the system. In the last section we
explicit the concluding remarks.

\section{A brief overview of gravity coupled to a scalar field}

In this section the work presented by Gremm is briefly revisited \cite{gremm}, where
4-dimensional gravity arises on a thick domain wall in {AdS} space. We
start with the action in 5-dimensional gravity coupled to one real scalar
field, which is given by%
\begin{equation}
\mathcal{S}=\int d^{5}x\sqrt{\left\vert g\right\vert }\left[ -\frac{R}{4}+%
\frac{1}{2}g^{ab}\nabla_a \phi\nabla_b \phi-V(\phi )\right] ,
\end{equation}
\noindent where $4\pi G=1$, with the field and the space-time variables
being dimensionless, $R$ denotes the scalar curvature, and the scalar field $%
\phi$ depends only on the extra dimension. Furthermore, $V(\phi )$ is the
potential describing the model, $g=\det (g_{ab})$, and the metric is
represented by 
\begin{equation}
ds^{2}=g_{ab}dx^{a}dx^{b}=e^{2A(r)}\eta_{\mu \nu }dx^{\mu }dx^{\nu
}-dr^{2}\,,
\end{equation}
\noindent for $a,b=0,1,2,3,5$, where $r$ is the extra dimension, $\eta_{\mu
\nu }$ denotes the usual Minkowski metric components, and $e^{2A(r)}$ is the  warp
factor.

By denoting $B^{\prime }(r)=dB(r)/dr$, for any quantity $B$ depending just
upon the variable $r$, and using the Einstein equations $\mathcal{G}_{ab}=%
\mathcal{T}_{ab}$ and the Euler-Lagrange equation $\nabla \phi _{a}\nabla
\phi ^{a}+V^{\prime }(\phi) =0$ as well, the following equations of motion
read: 
\begin{eqnarray}
A^{\prime \prime }(r)+\frac{2}{3}\phi^{\prime 2}(r) &=&0,  \label{2.1} \\
A^{\prime 2}(r)-\frac{1}{6}\phi^{\prime 2}(r)+\frac{1}{3}V(\phi ) &=&0,
\label{2.2} \\
\phi^{\prime \prime }(r)+4\phi^{\prime }(r)A^{\prime }(r)-V^{\prime }(\phi)
&=&0.  \label{2.3}
\end{eqnarray}
Moreover, the above equations are equivalently written as 
\begin{eqnarray}
\phi^{\prime }(r) = \frac{1}{2}\frac{dW(\phi)}{d\phi}, \qquad A^{\prime
}(r)=-\frac13W(\phi)\,
\end{eqnarray}
whenever the potential $V(\phi)$ is provided by the superpotential $%
W(\phi)$ as \cite{De_Wolfe_PRD_2000}.
\begin{equation}
V(\phi )=\frac{1}{8}\left( \frac{dW(\phi )}{d\phi }\right) ^{2}-\frac{1}{3}
W^2(\phi)\,.  \label{2.4}
\end{equation}
\noindent 
Thus, it is straightforward to verify that the first-order equations 
\begin{eqnarray}
A^{\prime }(r) &=&-\frac{1}{3}W(\phi ),  \label{2.5} \\
\phi^{\prime }(r) &=&\frac{1}{2}\frac{dW(\phi )}{d\phi },  \label{2.6}
\end{eqnarray}
\noindent also solve Eqs. (\ref{2.1}), (\ref{2.2}) and (\ref{2.3}). In order
to find analytical solutions, the sine-Gordon model is employed \cite%
{skyrme1, skyrme2}, being defined by the following superpotential: 
\begin{equation}
W(\phi )=3\alpha \beta \sin \left( \sqrt{\frac{2}{3\alpha }}\phi \right) .
\label{2.7}
\end{equation}
By using the above equation in Eq. (\ref{2.4}), the potential yields 
\begin{equation}
\!\!\!V(\phi )=\frac{3\alpha \beta ^{2}}{8}\left[ (1-4\alpha
)\!-\!(1+4\alpha )\cos \left( \sqrt{\frac{8}{3\alpha }}\phi \right) \right] .
\label{2.8}
\end{equation}

Now the solutions of Eqs. (\ref{2.5}) and (\ref{2.6}) are straightforwardly
verified to be given by 
\begin{eqnarray}
A(r) &=&-\alpha \ln \left[ 2\cosh (\beta r)\right] ,  \label{2.9} \\
\phi (r) &=&\sqrt{6\alpha }\arctan \left[ \tanh \left( \frac{\beta r}{2}%
\right) \right] .  \label{2.10}
\end{eqnarray}

The field $\phi (r)$ and the warp factor $e^{2A(r)}$ are shown,  respectively, 
in Figs. 1 and 2. The field configuration in Fig. 1 is evinced to be
the so-called kink. Moreover, $e^{2A(r)}$ is centred on $r=0$. It is
important to remark that in the solutions (\ref{2.9}) and (\ref{2.10}) the  {%
AdS} curvature is related by the product $\alpha \beta $, whereas the
thickness of the domain wall is given by the parameter $\beta $. In
addition, the brane core is localized at $r = 0$, which is consonant to the
point of maximum variation of the scalar field, what composes the well-known
domain wall. The thick brane presents a thickness $\Delta$ where deviations
from the 4D Newton's law eventuate in these scales.  Current precision
torsion-balance experiments require that the extra dimension must 
satisfy the constraint $\Delta \lesssim 44\,\mu$m~\cite{Kapner:2006si},
whereas theoretical reasons thus imply that $\Delta\gtrsim\ell_{(5)} \approx
2.0 \times 10^{-19}\,$m. Indeed, since in the 5D scenario with $%
M_{(5)}\simeq M_{\mathrm{ew}}\simeq 1\,$TeV, to be considered, the electroweak scale corresponds to the length $%
\ell_{(5)}\simeq 2.0\times 10^{-19}$m. Besides, it was shown in Ref. \cite%
{hoff1} that the above experimental and theoretical  arguments impose physical
constraints for the parameter space $\alpha\beta$: 
\begin{equation}
1.0\times 10^{-19} \beta \lesssim \arccos{\hspace{-0.5mm}\mathrm{h}}\left(%
\frac{ 10^{13/\alpha}}{2}\right) \lesssim 2.2 \times 10^{-5}\beta.
\label{cvb}
\end{equation}
\begin{figure}[h]
\begin{center}
\includegraphics[width=8.7cm]{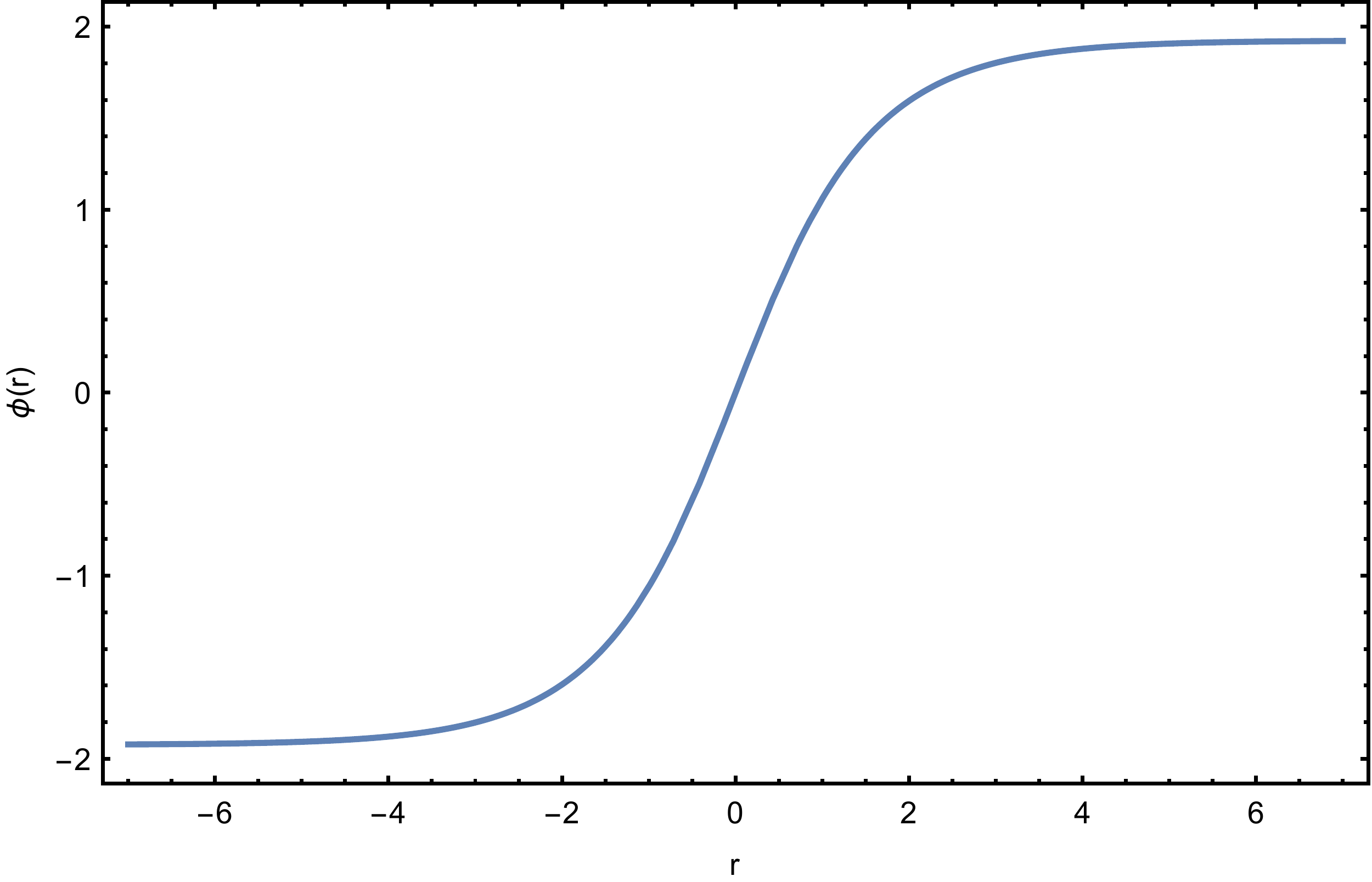}
\end{center}
\caption{Kink-like solution for $\protect\alpha=\protect\beta=1$.}
\label{fig1a:Solution}
\end{figure}
\begin{figure}[h!]
\begin{center}
\includegraphics[width=8.7cm]{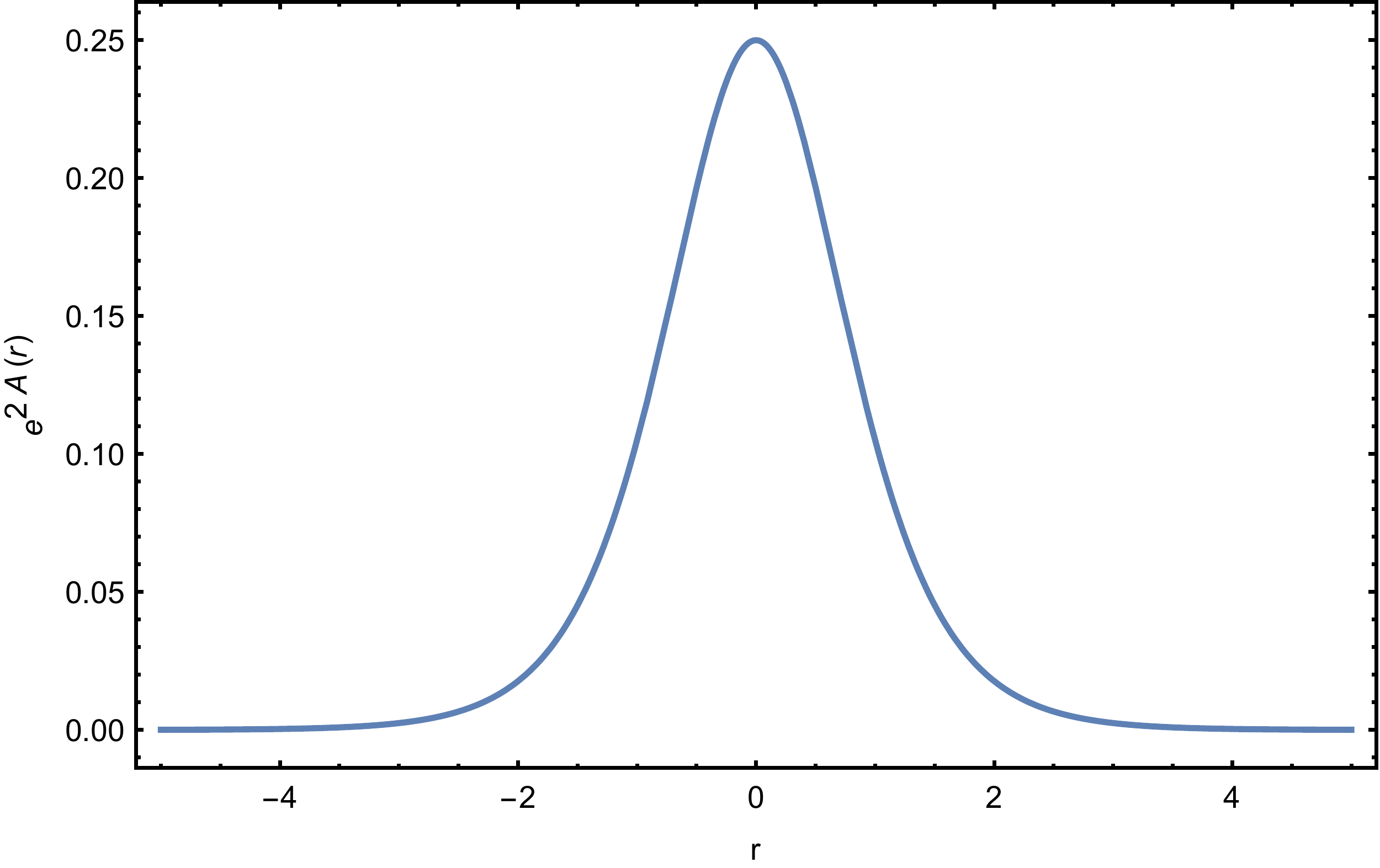}
\end{center}
\caption{Warp factor with $\protect\alpha=\protect\beta=1$.}
\label{fig1b:Solution}
\end{figure}

In the next section we shall postulate the configurational entropy in
brane-world scenarios. As an example, the sine-Gordon model described here
shall be explored.

\section{brane-world Configurational Entropy (BCE)}

As argued in the Introduction, Gleiser and Stamatopoulos (GS) \cite%
{gleiser-stamatopoulos} have recently proposed a detailed picture of the
so-called Configurational Entropy (CE) for the structure of localized
solutions in classical field theories. In this section, analogously to that
work, we formulate a configurational entropy measure in the functional
space, from the field configurations where brane-world scenarios can be
studied. Firstly, the framework shall be formally introduced and thereafter its
consequences are going to be explored. Hence, we discuss a prominent feature
of this theory.

To start, let us write the following Fourier transform%
\begin{equation}
\mathcal{F}[\omega ]=-\frac{1}{\sqrt{2\pi }}\int dr\;e^{2A(r)+i\omega r}%
\mathcal{L},  \label{3.1}
\end{equation}%
where $\mathcal{L}$ is the standard Lagrangian density and $e^{2A(r)}$
denotes the warp factor, as usual. For the sake of simplicity, define $%
\varepsilon (r):=-\mathcal{L}e^{2A(r)}$, named the warp density (WD). Thus,
using the Plancherel theorem it follows that%
\begin{equation}
\int d\omega \left\vert \mathcal{F}[\omega ]\right\vert ^{2}=\int
dr\left\vert \varepsilon (r)\right\vert ^{2}.  \label{3.1.1}
\end{equation}

Now the modal fraction is defined  by the following expression \cite{gleiser-stamatopoulos,
gleiser-sowinski, rafael-dutra-gleiser, rafael-roldao-dutra,
gleiser-sowinski2}: 
\begin{equation}
f(\omega )=\frac{\left\vert \mathcal{F}[\omega ]\right\vert ^{2}}{\int
d\omega \left\vert \mathcal{F}[\omega ]\right\vert ^{2}},  \label{3.2}
\end{equation}%
The modal fraction $f(\omega )$ measures the relative weight of each mode $%
\omega $. Analogously to the Shannon's information theory, the CE can be
described by the expression%
\begin{equation}
S_{c}[f]=-\int d\omega \tilde{f}(\omega )\ln [\tilde{f}(\omega )],
\label{3.3}
\end{equation}%
\noindent where $\tilde{f}(\omega ):=$ $f(\omega )/f_{\max }(\omega )$ is
defined as the normalized modal fraction, whereas the term  $f_{\max }(\omega
)$ denotes the maximum fraction. Thus, Eq. (\ref{3.1}) can be used to
generate the modal fraction, in order to obtain the entropic profile of
 thick brane solutions. It is important to remark that Eq. (\ref{3.1}) differs from
that one given by GS. In this framework we are including the warp effect in
the function $\mathcal{F}[\omega ]$, and consequently the framework  carries
information about the warped geometry.

Here, as a straightforward example, we shall calculate the entropic information for
the sine-Gordon model. By substituting Eq. (\ref{2.4}%
) and Eq. (\ref{2.6}) as well into the WD, and after straightforward manipulations,
it yields 
\begin{equation}
\varepsilon (r)=e^{2A}\left( \frac{1}{4}W_{\phi }^{2}-\frac{1}{3}%
W^{2}\right) .  \label{3.4}
\end{equation}
With the sine-Gordon model provided by Eq. (\ref{2.7}) and its respective
solutions, the above WD can be written in the following form: 
\begin{equation}
\!\!\!\!\!\varepsilon (r)\!=\!\frac{6\alpha \beta ^{2}\!\cosh^{\!-2}(\beta r)%
}{\left[ 1+\tanh ^{2}(\beta r)\right] ^{2}}\left[\sech^{4}(\beta
r)\!-\!6\alpha \tanh ^{2}(\beta r)\right] .  \label{3.5}
\end{equation}
Now, the Fourier transform of the WD is calculated, which gives the modal
fraction in Eq. (\ref{3.2}). In fact, we must determine%
\begin{equation}
\mathcal{F}[\omega ]=\frac{1}{\sqrt{2\pi }}\int dr\;e^{i\omega r}\varepsilon
(r),  \label{3.6}
\end{equation}
\noindent \noindent with $\varepsilon (r)$ given by Eq. (\ref{3.5}). After
exhaustive calculations, one finds%
\begin{equation}
\mathcal{F}[\omega ]=\frac{2^{1-2\alpha }\alpha \beta ^{2}}{\sqrt{2\pi }}%
\sum_{j=1}^{2}\left[ \frac{2(2+9\alpha )}{3}I_{1}^{(j)}-\frac{3\alpha }{4}%
I_{2}^{(j)}\right] ,  \label{3.7}
\end{equation}
\noindent where $I_{1}^{(j)}$ and $I_{2}^{(j)}$ are functions given by 
\begin{eqnarray}
\!\!\!I_{1}^{(j)} &=&\frac{1}{2\beta }\frac{\Gamma (\lambda _{j}+1)}{\Gamma
(\lambda _{j}+2)}{}_{2}\emph{G}_{1}[\gamma _{j},\lambda _{j}+1;\lambda
_{j}+2;-1]\,, \\
&&  \notag \\
\!\!\!I_{2}^{(1)} &=&4(\bar{I}_{2}^{(1)}+\tilde{I}_{2}^{(1)}),\quad \text{ }%
\!\!\!\!\!\!I_{2}^{(2)}=4(\bar{I}_{2}^{(2)}+\tilde{I}_{2}^{(2)})\,.
\label{3.8}
\end{eqnarray}
The above functions are defined as 
\begin{eqnarray}
\!\!\!\!\!\!\bar{I}_{2}^{(j)} &=&\frac{1}{2\beta }\frac{\Gamma (\bar{\lambda}%
_{1}+1)}{\Gamma (\bar{\lambda}_{1}+2)}{}_{2}\emph{G}_{1}[\bar{\gamma}_{1},%
\bar{\lambda}_{1}+1;\bar{\lambda}_{1}+2;-1], \\
&&  \notag \\
\!\!\!\!\!\!\tilde{I}_{2}^{(j)} &=&\frac{1}{2\beta }\frac{\Gamma (\tilde{%
\lambda}_{1}+1)}{\Gamma (\tilde{\lambda}_{1}+2)}{}_{2}\emph{G}_{1}[\tilde{%
\gamma}_{1},\tilde{\lambda}_{1}+1;\tilde{\lambda}_{1}+2;-1]\,,
\end{eqnarray}
where the above expressions $_{2}\emph{G}_{1}[\;\cdot\;,\;\cdot\;;\;\cdot\;;%
\;\cdot\;]$ stand for the well-known hypergeometric functions with%
\begin{eqnarray}
\gamma _{1} &=&\gamma _{2}=\bar{\gamma}_{m}=\tilde{\gamma}_{m}=2(\alpha +1),
\notag \\
\lambda _{1}&=&\alpha +i\omega /2\beta ,\qquad \lambda _{2}=\lambda
_{1}^{\ast }, \qquad \bar{\lambda}_{1} =\lambda _{1}+1,  \notag \\
\bar{\lambda}_{2}&=&\bar{\lambda}_{1}^{\ast },\qquad \tilde{\lambda}%
_{1}=\lambda _{1}^{\ast }-1,\qquad \tilde{\lambda}_{2}=\lambda _{1}-1\,, 
\notag
\end{eqnarray}
\noindent where the $\lambda^*$ denotes the complex conjugate of $\lambda$.
Now, in order to lead the modal fraction to a more compact form, Eq. (\ref%
{3.7}) can be rewritten as: 
\begin{equation}
\mathcal{F}[\omega ]=A_{0}\sum_{j,k=1}^{2}c_{k}I_{k}^{(j)},
\end{equation}
\noindent where the following notation is used:%
\begin{equation}
A_{0}=\frac{2^{1-2\alpha }\alpha \beta ^{2}}{\sqrt{2\pi }},\text{\;\;}c_{1}=%
\frac{2(2+9\alpha )}{3},\text{\;\; }c_{2}=-\frac{3\alpha }{4}.
\end{equation}
Thus, the modal fraction Eq. (\ref{3.1}) becomes 
\begin{equation}
f(\omega )=\frac{\displaystyle\sum_{j,k,m,n=1}^{2}c_{k}c_{n}^{\ast
}I_{k}^{(j)}I_{n}^{(m)\ast }}{\displaystyle\sum_{j,k,m,n=1}^{2}\int d\omega
c_{k}c_{n}^{\ast }I_{k}^{(j)}I_{n}^{(m)\ast }}.  \label{3.9}
\end{equation}
In Figs. 3 and 4 the modal fraction is depicted, for different values of the
parameter $\alpha$. Note that the maximum of the distributions are localized
around the zero mode $\omega =0$. Indeed, the maximum fraction is given by $%
f_{\max }=f_{\max }(0)$. 
\begin{figure}[h]
\begin{center}
\includegraphics[width=8.7cm]{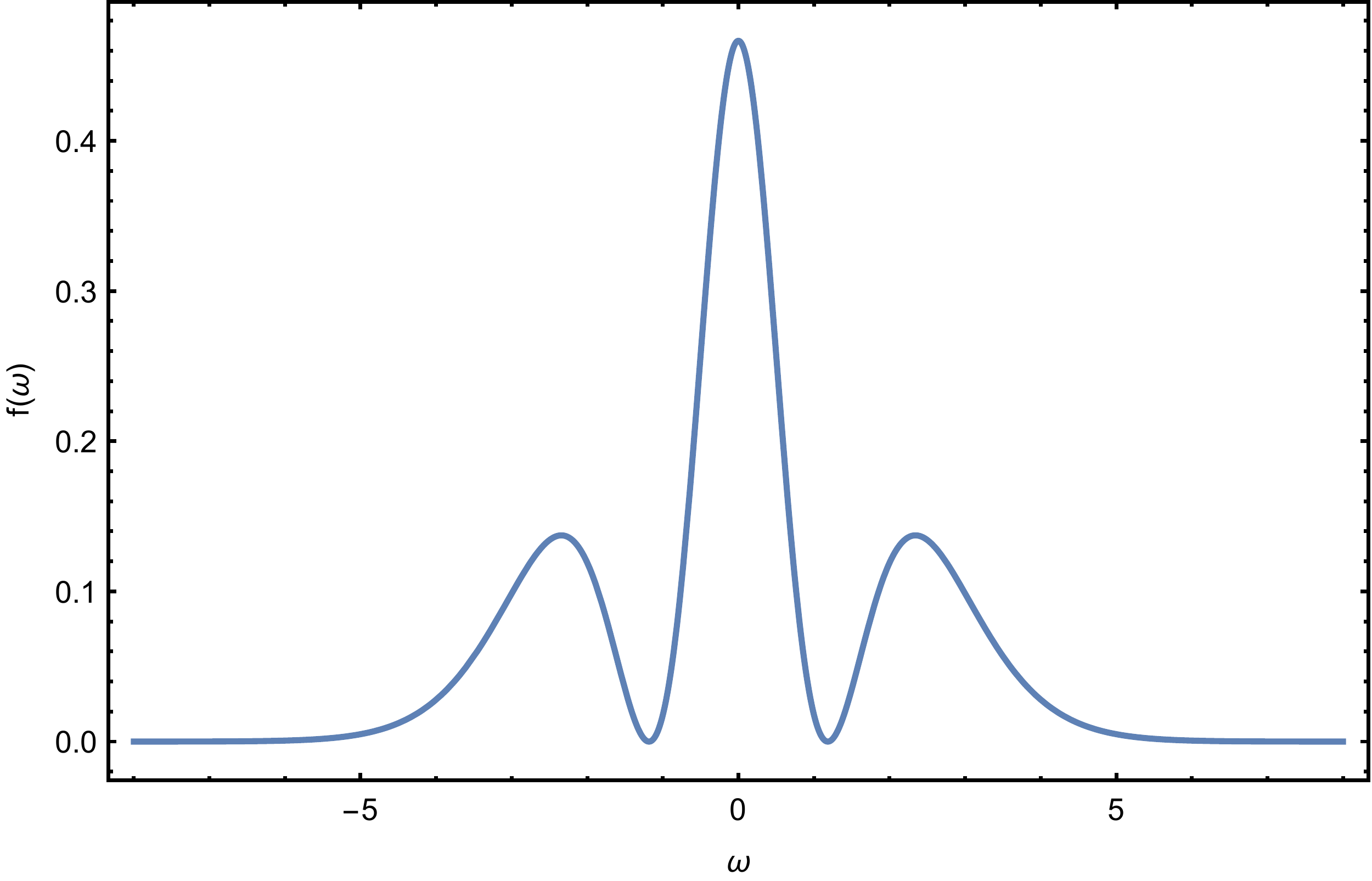}
\end{center}
\caption{Modal fractions for $\protect\alpha=\protect\beta=1$. The maximum
is at $\protect\omega=0$. }
\label{fig2a:Fractions}
\end{figure}
\begin{figure}[h]
\begin{center}
\includegraphics[width=8.7cm]{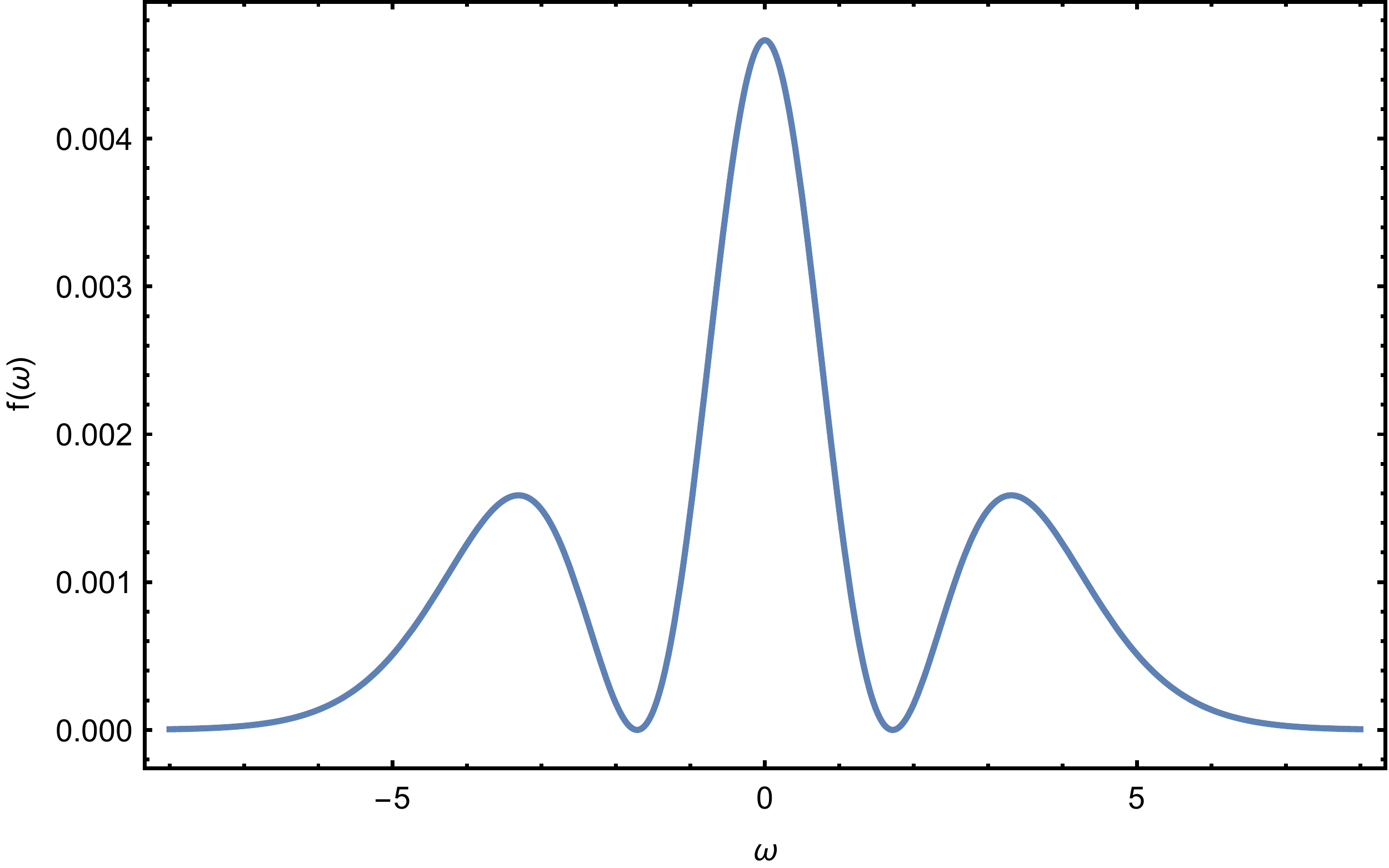}
\end{center}
\caption{Modal fractions for $\protect\alpha=2$ and $\protect\beta=1$. Note
that the maximum is also at $\protect\omega=0$. }
\label{fig2b:Fractions}
\end{figure}
By taking into account the modal fraction in Eq. (\ref{3.9}) and its maximum
contribution, Eq. (\ref{3.3}) can be now solved in order to obtain the brane
CE. In this case, due to the high complexity of integration, Eq. (\ref{3.3})
must be integrated numerically. The results are shown in Fig. 5, where the
BCE is plotted as a function of the parameter $\alpha $. In this case, the
rescaled parameter $\delta =\alpha \beta $ have been defined, which is
responsible by the {AdS} curvature. Moreover, the thickness of the wall is
fixed by taking $\beta =1,2,3$. As a consequence, several remarkable results
can be enumerated here. First, we have found that the entropic information
measure provides an accurate way to fix the best values of the AdS curvature. In fact, we demonstrate that the best ordering is  given by that ones with lower values of
the domain wall thickness. Second, at large values of $\alpha$, the brane-world CE yields the
configurational entropy $S_{c}=0$, showing a great organisational degree in
the structure of configuration of the system.

Finally, by using a recent approach presented by GS \cite{gleiser-stamatopoulos}%
, we have checked that the BCE is correlated to the energy of the system. The higher
[lower] the brane configurational entropy, the higher [lower] the energy of
the solutions. 
\begin{figure}[h]
\begin{center}
\includegraphics[width=8.7cm]{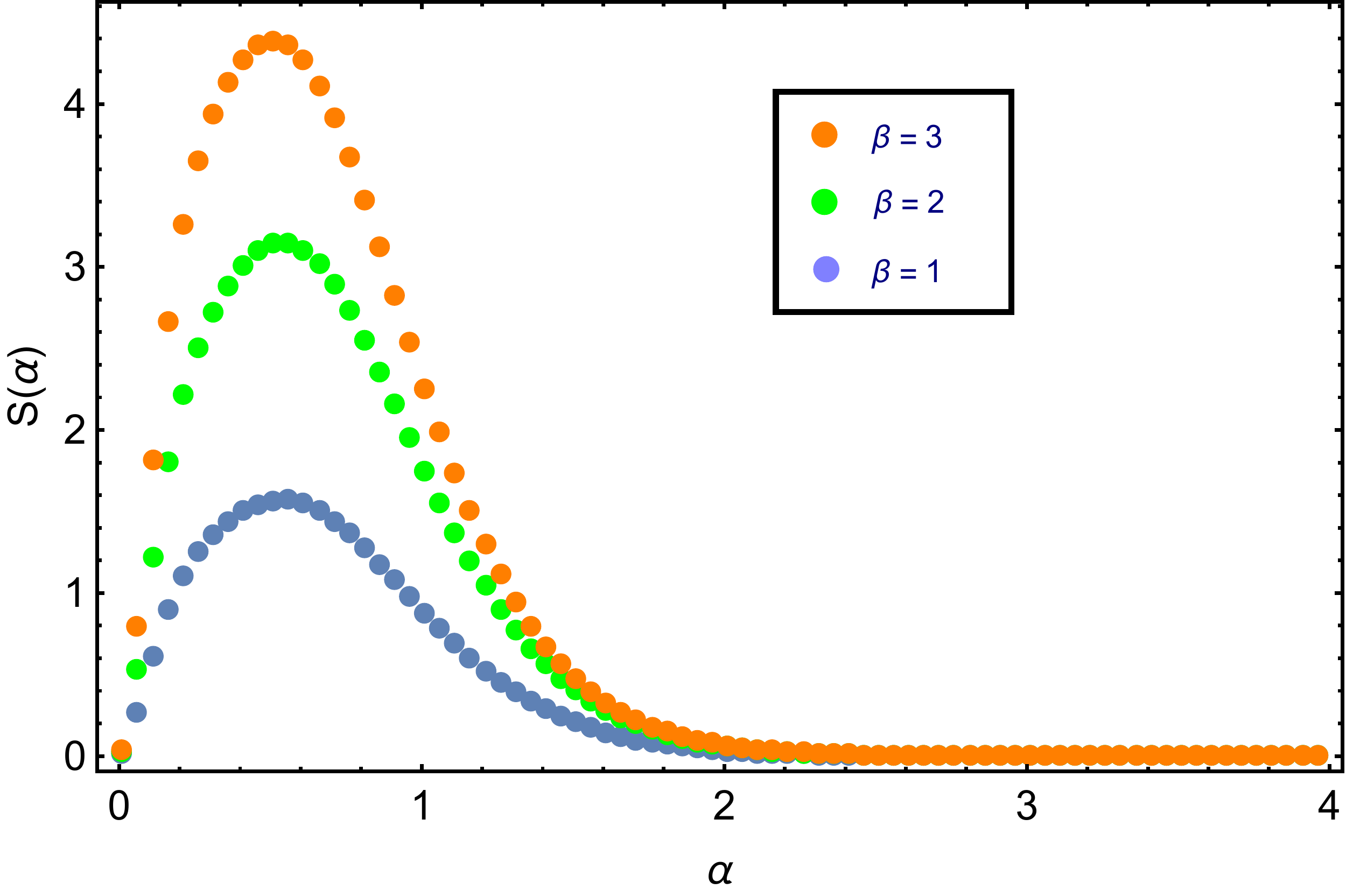}
\end{center}
\caption{The brane configurational entropy as a function of the parameter $%
\protect\alpha$.}
\label{fig3:Entropy}
\end{figure}
Moreover, it is straightforward to realize from Fig. \ref{fig3:Entropy} that the
higher the brane thickness $\beta$, the higher the respective value for the
brane configurational entropy as well.

In the next section we shall provide the consequences of the model above
studied and point forthcoming perspectives.

\section{Concluding Remarks and Outlook}

The entropic information has been here studied in brane-world models, with
emphasis on the sine-Gordon model, which has been chosen by its very
physical content and usefulness. Indeed, the sine-Gordon model parameters
provide the AdS bulk curvature and the  domain wall thickness as well. Hence
the brane-world entropic information for the sine-Gordon model has been
achieved, providing the most suitable values for the AdS curvature. In fact,
we proved that the higher the brane thickness $\beta$, the higher the
respective value for the brane configurational entropy. The brane-world
configurational entropy is moreover exerted to evince a higher
organisational degree in the structure of system configuration likewise, for
large values of one of the sine-Gordon model parameters. The Gleiser and
Stamatopoulos procedure was also used to acquire the correlation
between the the brane-world configurational entropy and the energy of the
system, withal.

Once developed the formalism of the brane-world configurational entropy and
the entropic information as well, we can further apply a procedure similar
to what has been studied in the previous sections to other thick brane-world
models. 
Indeed, an entire new family of models of the sine-Gordon type, starting
from the sine-Gordon model, including the double sine-Gordon, the triple
one, and so on, have been obtained in \cite{epjc}. Such models appear as
deformations of the starting sine-Gordon model, and as they present different topological sectors, it would be important to
probe them from the point of view of the brane-world configurational
entropy. Since the solutions of these deformed models can be constructed
explicitly from the topological defects of the sine-Gordon model itself, we
expect to study in particular the double sine-Gordon model in a brane-world scenario with
a single extra dimension of infinite extent, as in this framework a stable
gravity scenario has been shown to be allowable \cite{epjc}. Other
interesting brane-world models that are beyond the scope here, as for
instance the Bloch branes \cite{ThickBrane2}, the cyclically deformed topological defects that generate domain walls \cite{Bernardini:2012bh}, and the asymmetric sine-Gordon
model as well \cite{hoff,Bazeia:2013usa}, can be also investigated from the
point of view of the entropic information and the brane-world
configurational entropy. \acknowledgments
RACC thanks to UFABC and CAPES for financial support. RdR thanks to SISSA
for the hospitality and to CNPq Grants No. 303027/2012-6 and No.
473326/2013-2 for partial financial support. RdR is also supported in part
by the CAPES Proc. 10942/13-0.

\end{document}